# Identifying field-tunable surface resonance states on black phosphorus

*Dongming Zhao, Byeongin Lee, Junho Bang, Claudia Felser, Jian-Feng Ge\*, Doohee Cho\**

Dongming Zhao

Max Planck Institute for Chemical Physics of Solids, 01187 Dresden, Germany

Byeongin Lee, Junho Bang

Department of Physics, Yonsei University, Seoul 03722, Republic of Korea

Claudia Felser

Max Planck Institute for Chemical Physics of Solids, 01187 Dresden, Germany

Jian-Feng Ge

Max Planck Institute for Chemical Physics of Solids, 01187 Dresden, Germany

E-mail: Jianfeng.Ge@cpfs.mpg.de

Doohee Cho

Department of Physics, Yonsei University, Seoul 03722, Republic of Korea

E-mail: doohee cho@yonsei.ac.kr

Funding: B. L., J. B., and D. C. were supported by the National Research Foundation of Korea (NRF) grant funded by the Korea government (No. RS-2023-00251265, RS-2024-00337267, and RS-2024-00442483) and the Industry-Academy joint research program between Samsung Electronics and Yonsei University.

**Abstract:** Surface resonance states are electronic states localized near the surface while remaining hybridized with bulk bands. These states can strongly modify the electric-field response of semiconductors. Here, we demonstrate using scanning tunneling spectroscopy that on black phosphorus, surface resonance states near the valence-band edge dominate the screening of a strong external electric field. We observe in the tunneling conductance spectrum a pronounced dip with an energy continuously tunable by the local electric field in the tunneling junction. Meanwhile, we also notice that the bulk band edges remain effectively pinned, indicating efficient surface screening and suppression of bulk band bending. We interpret the conductance dip as the consequence of a field-dependent tunneling barrier: as the external electric field drives the surface resonance band into the band gap, the coupling between the surface resonance states and the bulk states is suppressed, leading to a reduced tunneling probability. Our simplified model based on this mechanism reproduces our main experimental findings. Our results highlight surface-localized states as a critical component in the electrostatic response of semiconductors, which must be taken into consideration in the design and operation of nanoscale semiconductor devices. (187 words)

# 1. Introduction

Semiconductors exhibit rich and highly tunable electronic responses under external electric fields, which play a central role in electronic, optoelectronic, and nanoscale device applications. Near surfaces and interfaces, electric fields can substantially reshape the local electrostatic potential, leading to charge redistribution and modified electronic spectra. Electric-field control is widely exploited, for example, to induce carrier accumulation and depletion in field-effect devices [1–2], and to engineer band profiles in semiconductor heterostructures and quantum-confined systems [3–4]. Understanding how electronic states near semiconductor surfaces respond to strong and spatially confined electric fields is of importance for nanoscale semiconductor devices, as electrostatic control, screening, and field-driven charge redistribution critically determine device operation and functionality.

Such field-induced effects, when probed at the atomic scale, become particularly pronounced, as strong and spatially confined electric fields can be generated locally. In scanning tunneling microscopy (STM), the bias voltage applied between a metallic tip and a semiconductor surface produces a highly localized electric field in the tunneling junction. This strong field (~ 1 V/nm) penetrates the semiconductor and modifies the near-surface electrostatic potential. This phenomenon, known as tip-induced band bending (TIBB) [5–9], is characterized by a depth-dependent rigid shift of the conduction and valence bands with respect to the bulk Fermi level near the surface. As a result, the energies of near-surface electronic states are also shifted relative to the bulk bands, strongly affecting the local electronic behavior.

For instance, surface resonance states can accommodate charge accumulation at the surface and therefore strongly modify the electrostatic response of semiconductors [10]. Unlike genuine in-gap surface states that are decoupled from the bulk states, surface resonance states retain a certain degree of coupling with bulk bands as well as finite spectral weight near the surface, enabling them to simultaneously participate in charge screening and tunneling processes [11–12]. In contrast to the conventional picture of TIBB, where both bulk band edges and other spectral features acquire the same, constant energy shift as the local electric field, the presence of surface resonance states can efficiently screen the bulk, resulting in pronounced spectroscopic signatures without a detectable shift of the bulk band edges [13–19].

Despite extensive studies of the TIBB phenomenon [5–9,16–27], it remains largely unexplored how semiconductors hosting surface resonance states react to strong local electric fields. In this work, we investigate the surface resonance states on black phosphorus (BP), which give rise to a pronounced dip in the tunneling conductance spectroscopy. Driven by TIBB, the dip shifts accordingly with the local electric field in the tunneling junction, whereas the bulk conduction and valence band edges remain unaffected by the field. We demonstrate that surface-localized electronic states can dominate the electrostatic response and provide a tuning knob for dielectric screening.

## 2. Results and Discussion

We performed STM measurements at a base temperature of 4.2 K on the surface of commercial p-type BP single crystals (HQ graphene). The cleaved BP (001) surface exhibits a characteristic puckered honeycomb lattice structure with two distinct directions (inset of **Figure 1a**), namely the zigzag (ZZ) and armchair (AC) [16–18,27]. As illustrated in **Figure 1a**, the typical topographic image of BP (001) shows an atomically flat clean surface region free of defects, where the stripe pattern corresponds to the alternating zigzag chains with different apparent heights.

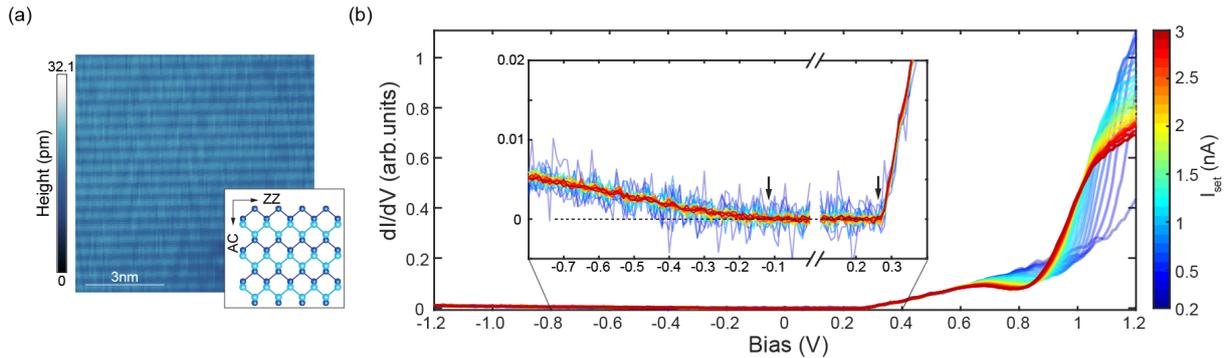

**Figure 1.** Topography and spectroscopy on the BP surface. (a) Typical STM image of the BP (001) surface. Inset shows the crystal structure, with brighter and darker spheres represent surface and subsurface phosphorus atoms, respectively. Zigzag (ZZ) and armchair (AC) directions are marked. Setup conditions: $V_B$ = 1.2 V, $I_{set}$ = 0.1 nA. (b) Normalized d$I$/d$V$ spectra taken at different tunneling setup current as indicated by the color bar. Inset shows the zoom-in view of the spectra near the conduction band minimum and valence band maximum (indicated by two arrows), respectively.

On this BP surface, we acquired a series of spectra at the same location while systematically varying the setup current. Changing the setup current effectively modifies the tip-sample distance and hence the strength of the electric field in the tunneling junction [28]. The overall differential conductance ($g$ = d$I$/d$V$) spectra (**Figure 1b**) show an apparent gap of ~ 0.34 eV (see Supplementary Material for details), in agreement with previous STM reports [16]. Our main observation here is a pronounced dip feature, centered around the bias of ~ 0.9 V. By extracting the dip position from the corresponding peak in the second derivative d$^2g$/d$V^2$ spectra, we find a substantial shift of the dip from a bias voltage of 1.10 V to 0.85 V as the setup current is increased from 0.1 nA to 3.0 nA. Although the differential conductance spectroscopy commonly reflects the density of states, such a pronounced and continuously tunable spectral feature is clearly incompatible with intrinsic bulk electronic states. We can also exclude impurity-related origins for the observed conductance dip, as no visible defects exist within 10 nm around the location of the acquired spectra. This rules out the influence of impurities, which could accommodate charge and induce localized states [11–12].

In addition, we notice, in the same set of spectra, that the edges of both the conduction band and the valence band remain essentially unchanged (inset of **Figure 1b**), while the conductance dip exhibits a strong and systematic dependence on the tunneling current. This behavior is unusual for semiconductors under TIBB, where bulk band edges are generally expected to shift with the local electrostatic potential [5–12] (dashed curves in **Figure 2a**). Such field-insensitive behavior of band edges was interpreted in previous reports [16,30] as strong pinning by surface-localized electronic states, which efficiently screen the tip-induced potential (solid curves in **Figure 2a**). The coexistence of a strongly field-dependent conductance dip and field-insensitive bulk band edges suggests that states localized on the surface play a dominant role in the determination of the tunneling response under local electric field [31]. Localized states on the BP surface were previously reported in photoemission experiments [32–34], identified as surface resonance states in the vicinity of the valence-band maximum. Despite their hybridization with the bulk valence bands, these states carry substantial near-surface spectral weight [10–12], making their energy strongly susceptible to the tip-induced electrostatic potential.

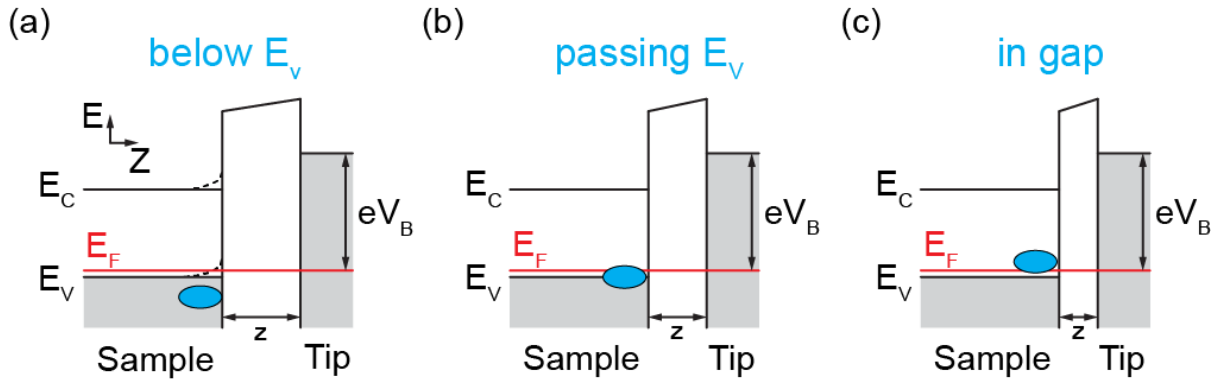

**Figure 2.** Schematic illustration of field-tunable surface resonance states with decreasing tip-sample separation $z$. (a) Energy diagram for a low external field, where surface resonance states (blue) stay coupled to bulk states (gray). Charge accumulation in the surface resonance states screens the electric field, pinning the bulk band edges (solid lines), in contrast to the tip-induced bending of bulk bands (dashed lines). $E_c$ and $E_v$ denote the conduction- and valence-band edges,

and $E_F$ the Fermi level (red line), respectively. (b–c) Increasing the local electric field by reducing the tip–sample separation $z$ shifts the surface resonance states upward, passing $E_v$ (b) and eventually enter the bulk energy gap (c). The overlap between the surface resonance states and the bulk band decreases during this process.

We attribute the conductance dip, therefore, to the surface resonance states near the valence-band edge of BP. The dip arises from a field-driven suppression of the coupling between the bulk bands and surface resonance states. When these surface resonance states reside within the valence-band continuum, they remain strongly hybridized with the bulk states (**Figure 2a**) [35]. As the electric field in the tunneling junction increases, the surface resonance states are shifted to a higher energy and eventually driven into the energy gap (**Figure 2b–c**) [9]. In such cases, their hybridization with the bulk continuum is strongly reduced, and thus electrons tunneling into these surface-localized states can no longer relax into the bulk states. As a result, the tunneling probability of this path into the bulk is suppressed [35–36], which in turn results in a dip in the conductance spectrum as the contribution from the surface resonance states decreases (see Supplementary Material for details). With an increasing setup current, the tip-sample distance is reduced, and therefore the dip position shifts accordingly toward a lower bias to reach the same electric field for the suppressed tunneling. The observed conductance suppression shares a similar mechanism with negative differential conductance observed by STM experiments in the sense of field-driven modification of tunneling probability [37–39], although the latter is based on resonant tunneling through *discrete* localized energy levels.

To verify the above physics underlying the conductance dip, we construct a simplified model that describes how the tunneling probability evolves as surface resonance states are tuned by an external electric field (see Supplementary Material for details). In our model, the low-energy bulk bands of BP are described by simple parabolic dispersions [40], while the surface resonance states are phenomenologically approximated by a Gaussian spectral feature located near the valence-band edge [10–12,33–35]. Under the tip-induced field, the center of the Gaussian peak is driven upward in energy with increasing electric field (strong/weak TIBB drives it faster/slower) [9]. The tunneling conductance is then calculated, including the effective tunneling probability that accounts for the coupling between surface resonance and bulk states [28–29]. Our simulation results in **Figure 3** show that, when the surface resonance states are driven into the energy gap, the suppressed relaxation into bulk states results in a reduction of the tunneling conductance [35–36]. The calculated differential conductance $g$ exhibits a pronounced dip (**Figure 3a**), and a corresponding peak feature exists in the $d^2g/dV^2$ spectra (**Figure 3b**). These features closely resemble the experimental observations shown in **Figure 3c–d**. From our simulation, we find that the dip voltage $V_{dip}$ (extracted from the peak position in the $d^2g/dV^2$ spectra) corresponds to the bias at which the peak of the surface resonance spectral weight is aligned with the gap edge. Our simulation shows a qualitative agreement with the experimental data, where $V_{dip}$ is shifted to a lower value for a higher setup current (stronger TIBB).

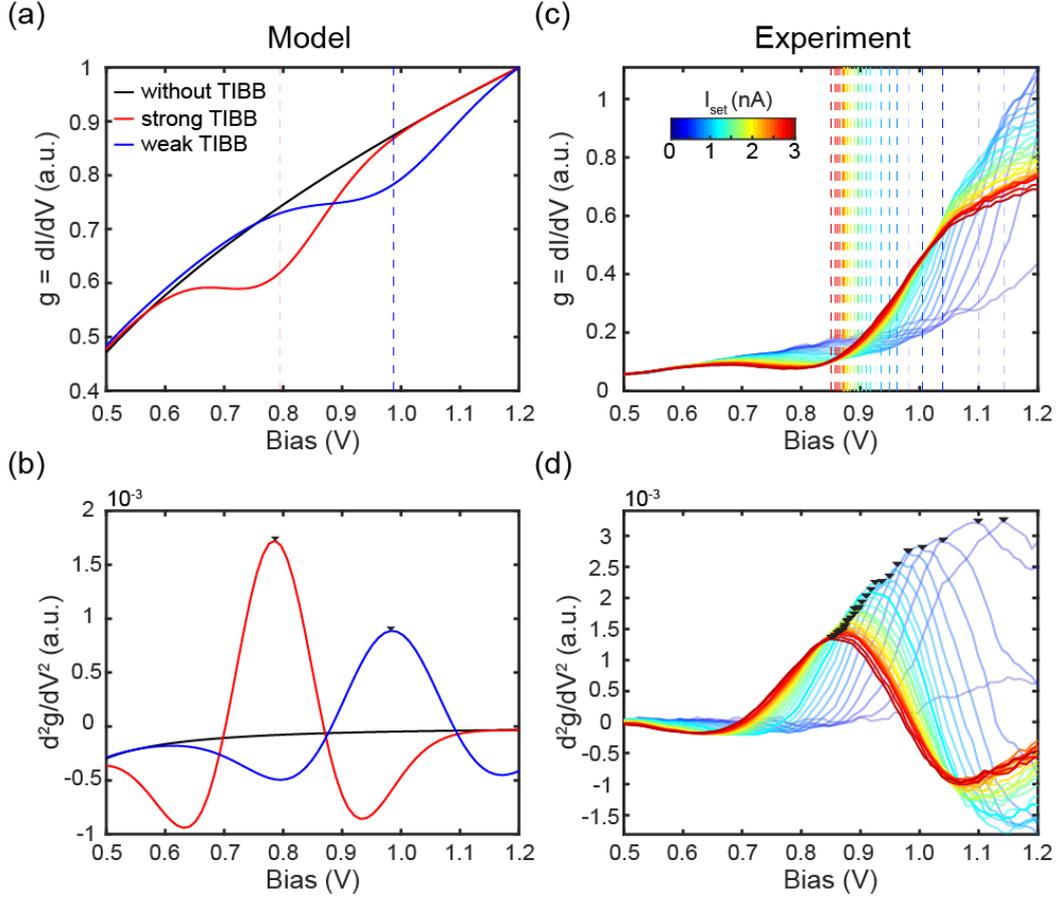

**Figure 3.** Comparison between the simplified model and experimental tunneling spectra. (a–b) Calculated differential conductance spectra ($g$=d$I$/d$V$) and second derivative of conductance spectra ($d^2g/dV^2$) without TIBB (black curves), with strong TIBB (red curves), and with weak TIBB (blue curves) and, respectively. (c–d) Experimental differential conductance spectra cropped from Figure 1b and corresponding $d^2g/dV^2$ spectra for different setup current, as indicated by the color bar. The triangles in (d) mark the peaks in $d^2g/dV^2$, which define the dip voltage $V_{dip}$. Dashed lines in (c) indicate the corresponding $V_{dip}$ positions in the $g(V)$ spectra.

Next, we analyzed how $V_{dip}$ depends on tunneling junction parameters. **Figure 4** shows that $V_{dip}$ varies monotonically with the tip–sample separation $z$ (as a reference, we set $z = z_0$ for $I_{set} = 1.0$ nA). For small $z$, $V_{dip}$ has an approximately linear dependence on $z$, with a slope corresponding to a constant electric field of 1.19 V/nm, but deviates from this linear behavior for $z > z_0$. Because the tip-sample separation $z$ is related to the effective tunnel barrier $\varphi$ by $\sqrt{\varphi} \propto -\mathrm{d}\ln I/\mathrm{d}z$ [28–29], we can examine the effective tunneling barrier $\varphi$ by plotting $\ln(I_{set})$ against $z$. We find in **Figure 4** that the $\ln(I_{set})$–$z$ data shows two regimes with distinct slopes as the tip approaches the surface, indicating a change in the effective tunneling barrier $\varphi$. This slope change also occurs at $z = z_0$, coinciding with the onset of nonlinearity of $V_{dip}(z)$. This coincidence indicates that tuning the energy of the surface resonance states can make a quantitative change in apparent barrier height and thereby in the screening behavior. For $z < z_0$, at the setup bias of 1.2 V, the complete depletion of the surface-resonance carriers alone is not enough to screen the tip electric field; the surface resonance states must also experience a $z$-linear band bending deeper into the band gap, as no bulk states can provide more carrier depletion path. This scenario is in principle the same as an intrinsic semiconductor. However, for $z > z_0$, the depletion of surface-resonance carrier is not complete at the setup bias of 1.2 V. A significant overlap of bulk and surface-resonance states restores the screening behavior of a doped semiconductor, allowing carriers to escape to the bulk. Therefore, with such a transition between an intrinsic to a doped semiconducting surface, we can see a quantitative change in the apparent barrier height. In addition, the nonlinear relationship between $V_{dip}$ and $z$ suggests that $V_{dip}$ corresponds to the point where surface-resonance carriers deplete fastest, and that the depletion rate of the carriers (and thus the TIBB strength) is reduced with decreasing field (see Supplementary Material for a detailed discussion), which requires a more complicated simulation beyond our model with a constant TIBB strength.

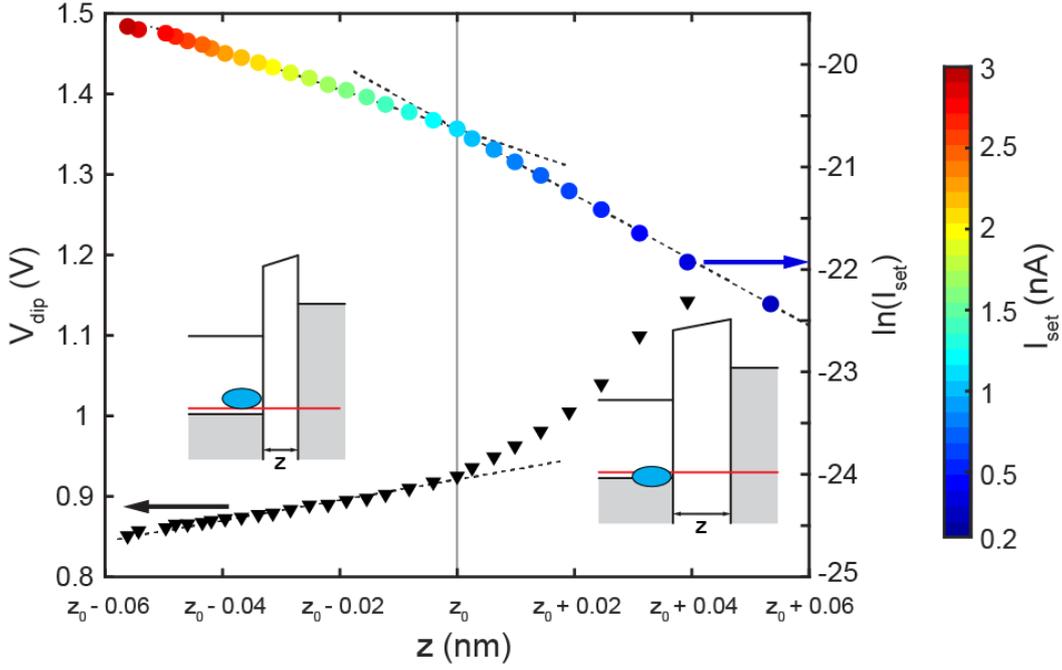

**Figure 4.** Comparison of the conductance dip position $V_{dip}$ (left axis, triangles) and the effective tunneling barrier (right axis, circles showing $\ln(I_{set})$) when varying the tip–sample separation z (with $z = z_0$ for $I_{set} = 1$ nA). Dashed lines show the approximated linear dependence. Gray line indicates the position $z_0$ where $\ln(I_{set}) - z$ data change slopes and $V_{dip} - z$ data deviate from linear dependence, dividing two regimes where surface resonance states start entering the bulk energy gap (right) and reside completely in the gap (left), as illustrated by the schematics.

## 3. Conclusion

In conclusion, we have demonstrated that surface resonance states on BP can dominate the tunneling response under a tip-induced electric field. By controlling this field, surface resonance states near the valence-band edge are driven into the bulk energy gap, leading to a redistribution of tunneling probability and a suppression of the tunneling conductance, while the bulk band edges remain effectively pinned due to efficient surface screening. A simplified model captures the essential experimental features and establishes a direct connection between electric-field-driven evolution of surface resonance states and the emergence of the conductance dip.

Our results emphasize in general that even on clean surfaces, surface resonance states can influence electronic transport in a manner comparable to defects. Similar to defects that dramatically alter device characteristics, surface resonance states can strongly reshape electrostatic screening under local electric fields, with direct implications for nanoscale semiconductor devices. Given the ubiquity of surface resonance states at real surfaces and interfaces [12], their influence on the properties of the charge carriers should be carefully considered in the design of semiconductor devices.

**Acknowledgements**

We thank M. Yao for valuable discussions.

**Data Availability Statement**

The data that support the findings of this article are not publicly available upon publication because it is not technically feasible and/or the cost of preparing, depositing, and hosting the data would be prohibitive within the terms of this research project. The data are available from the authors upon reasonable request.

# Supporting Information

**Identifying field-tunable surface resonance states on black phosphorus**

*Dongming Zhao, Byeongin Lee, Junho Bang, Claudia Felser, Jian-Feng Ge\*, Doohee Cho\**

**1. Sample preparation and STM measurement**

Commercial single crystals of black phosphorus (HQ-graphene) were used for STM measurement. The bulk crystals are cleaved at room temperature in the ultra-high vacuum with a base pressure of $1\times10^{-10}$ Torr. The samples were transferred to the precooled STM head. We performed STM measurements using a commercial low-temperature STM (USM1200LL, Unisoku Co., Ltd) at 4.2 K. We used mechanically sharpened Pt/Ir tips, whose condition was verified on the Au (111) surface. STM images were acquired in constant current mode, while differential conductance spectra were obtained using a standard lock-in technique with a modulation frequency of 913 Hz and a peak-to-peak amplitude of 10mV.

**2. Band gap determination**

The band gap value was determined by extracting the conduction band minimum (CBM) and valence band maximum (VBM) from the spectra in Figure 1b which have good signal-to-noise ratio (12 spectra with tunneling current setpoints larger than 1.5 nA). The CBM was extracted using the protocol following B. Kiraly et.al (Figure S1)[1]. First, the d$I$/d$V$ spectra were vertically offset by a factor of 1.1 times the absolute value of the minimum conductance to ensure no data point was negative. Second, the logarithm of the data was taken to extract the noise floor of each measurement. The in-gap conductance $C_g$ was determined by the average between 0.0 V and 0.2 V, and the standard deviation is $s_g$ (pink and blue lines). The CBM were then determined by the energy position where $\ln(dI/dV) > C_g + s_g$ (blue arrow). The VBM, however, require different method to extract due to its very low spectral intensity (the same method with CBM would give unreasonably low VBM value, resulting in unreasonably large band gap). We use a linear fit for the logarithm conductance between -0.4 V and -0.2V with linear function (red dashed line) and determined the VBM as the intersection of this linear function and $C_g$ (red arrow). The band gap was then calculated to be $0.34 \pm 0.07$ eV.

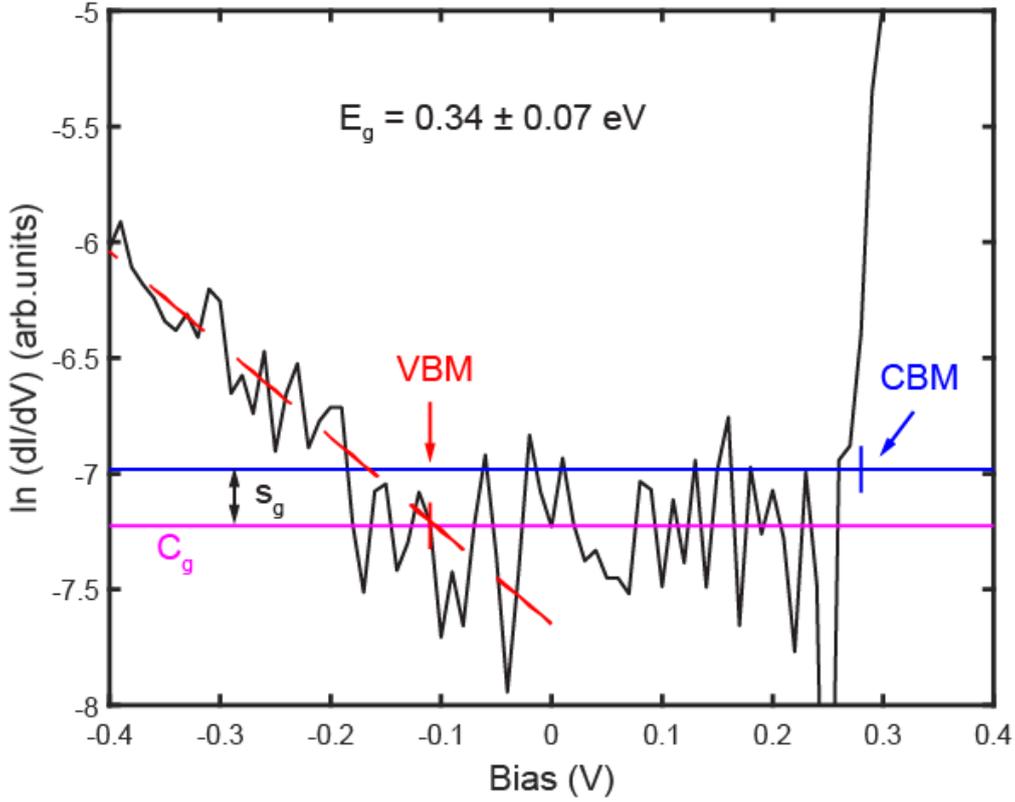

**Figure S1.** Logarithm of a d*I*/d*V* spectrum on BP. The horizontal lines indicate the average conductance $C_g$ in the band gap (pink) and the $C_g + s_g$ (red) levels for band determination. The dashed red line is the linear fit for the valence band data. The blue and red arrows point the CBM and VBM, respectively.

## 3. d*I*/d*V* signal with field-driven states

We start from the steady-state tunneling model for STM [2–3]

$$I \propto \int_0^{eV} \rho_S(E_F + \varepsilon) T(z, E_F + \varepsilon, V) d\varepsilon,$$

where *I* is the tunneling current, *V* the tunneling bias, $\rho_S$ the sample density of states, $E_F$ the energy of the Fermi level, *z* the tip-sample distance, and *T* the tunneling probability. Assumptions of low temperature and a constant tip density of states have been made. Following from previous work [4], when a trapezoidal tunnelling barrier and the WKB approximation is applied, the *T* is estimated to be energy-independent and only corresponds to the tip-sample distance *z*. At a fixed current setpoint (fixed *z*), we can omit the effect of *T*, and the d*I*/d*V* spectra then give the information of $\rho_S$

$$\frac{dI}{dV} \propto \rho_S(E_F + eV).$$

The main assumption in our model is that, for field-driven states like the surface resonance states in our case, $\rho_S$ also depends on the applied bias $V$

$$\rho_S(E,V) = \rho_B(E) + \rho_{SR}(E,V).$$

Here, $E$ is the energy, $\rho_B(E)$ is the field-independent bulk states, and $\rho_{SR}(E,V)$ is the field-dependent surface resonance states. In this case the corresponding d$I$/d$V$ becomes

$$\frac{dI}{dV} \propto \rho_B(E_F + eV, V) + \rho_{SR}(E_F + eV, V) + \int_0^{eV} \frac{\partial \rho_{SR}}{\partial V}(E_F + \varepsilon, V)d\varepsilon$$

$$= \rho_S(E_F + eV, V) + \int_0^{eV} \frac{\partial \rho_{SR}}{\partial V}(E_F + \varepsilon, V)d\varepsilon.$$

The first term is just the sample density of states. The second term is an additional part originate from the movement of the field-dependent states with respect to the bias.

To consider a simple and explicit case, we assume that this state can be described as a moving Gaussian peak

$$\rho_{SR}(E_F + \varepsilon, V) \propto e^{-\frac{(\varepsilon + E_P - sV)^2}{2c^2}}.$$

The peak position is set to be $E_F - E_P$ without any electric field (at zero bias), and shifts toward higher energy with increasing bias at a rate of $s$ eV per volt. $c$ is a constant.
The second term of d$I$/d$V$ is then

$$\int_0^{eV} \frac{\partial \rho_{SR}}{\partial V}(E_F + \varepsilon, V)d\varepsilon \propto s\left(e^{-\frac{(E_P - sV)^2}{2c^2}} - e^{-\frac{(eV + E_P - sV)^2}{2c^2}}\right)$$

$$= se^{-\frac{(E_P - sV)^2}{2c^2}}\left(1 - e^{-\frac{(eV)^2 + 2eV(E_P - sV)}{2c^2}}\right).$$

Thus, when the states are driven upwards in energy with tunneling bias at a small rate $s$ (< 0.5 eV per volt), the d$I$/d$V$ signal will have an additional positive part at positive applied bias.
In the case of our experiment, where the moving surface resonance states are driven into the energy gap, we need to take the effect of tunneling probability $T$. Qualitatively, when the surface resonance states are driven into the gap, their overlap with the bulk states decreases, leading to a suppression on the tunneling probability $T$[5–6].
When the effect of $T$ is taken into consideration, the d$I$/d$V$ becomes

$$\frac{dI}{dV} \propto \rho_S(E_F + eV, V)T(z, E_F + eV, V) + \int_0^{eV} \frac{\partial \rho_{SR}}{\partial V}(E_F + \varepsilon, V)T(z, E_F + \varepsilon, V)d\varepsilon$$

$$+ \int_0^{eV} \rho_{SR}(E_F + \varepsilon, V)\frac{\partial T}{\partial V}(z, E_F + \varepsilon, V)d\varepsilon.$$

The first term just gives $\rho_S(E_F + eV, V)$ at fixed $z$. Note that for a small bias, the surface resonance states stay below the Femi level, so $\rho_{SR}$ is zero while its derivative $\partial \rho_{SR}/\partial V$ is finite. For a larger bias when the surface resonance states are in the bulk gap, $T$ is strongly suppressed in the gap, while $\partial T/\partial V$ barely changes in the integral range of 0 to e$V$. Thus, the integral of the third term can be neglected. Comparing the second term with the integral in d$I$/d$V$ without the effect of $T$ (i.e., $T = 1$)

$$\int_0^{eV} \frac{\partial \rho_{SR}}{\partial V}(E_F + \varepsilon, V)\mathrm{d}\varepsilon,$$

the suppression of the $T$ ($T < 1$) for the surface resonance states in the gap causes the integral of

$$\int_0^{eV} \frac{\partial \rho_{SR}}{\partial V}(E_F + \varepsilon, V)T(z, E_F + \varepsilon, V)\mathrm{d}\varepsilon$$

to decrease, which in turn results in a conductance dip in the d$I$/d$V$ spectra.

A more quantitative treatment requires self-consistent Poisson–Schrödinger calculations including hybridized surface bands.

## 4. Simplified model

Black phosphorus (BP) is a predominantly p-type semiconductor, and its low-energy electronic band structure can be approximately modeled as simple parabolic dispersions for both conduction and valence bands[7]. Here in our experiments, the conductance dip lies in the conduction band. So, we only modeled the parabolic conduction bulk band dispersion (Figure S2).

The detailed dispersion of surface resonance states is more complicated in most cases. Here in our model, we described the surface resonance states phenomenologically by a Gaussian spectral peak feature with a full width at half maximum (FWHM) of 50meV whose energy position located at -200meV near the valence-band edge when no electric field is presented (Figure S2)[8–10]. It is shown later that this simplification is sufficient to capture the effect of surface resonance states on the total differential conductance spectra.

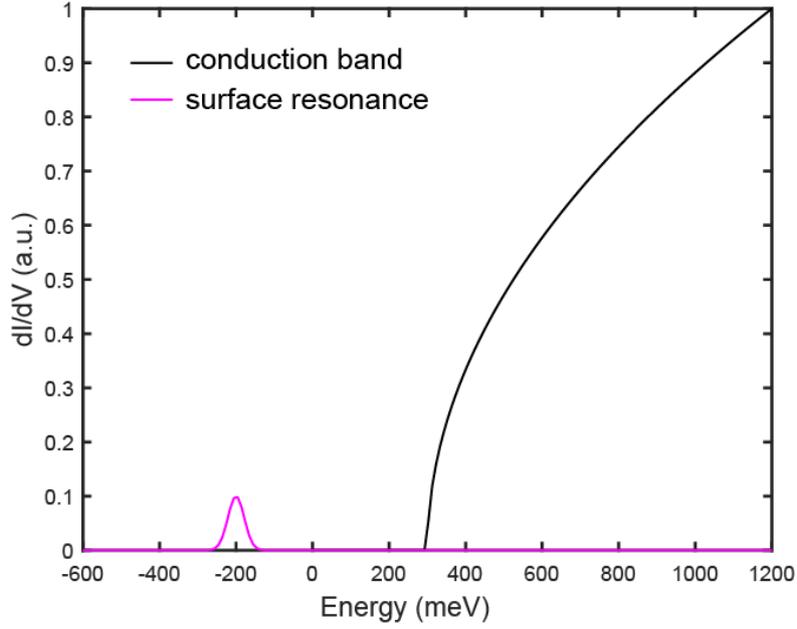

**Figure S2.** Modeled conduction band and surface resonance states of BP.

Under the influence of tip induced band bending (TIBB), while the conduction band is fixed due to the pinning effect, the surface resonance states are driven upwards to higher energy. This is described by moving the center of the Gaussian peak upward in energy when the tunneling bias is increasing, which corresponds to increasing the electric fields in the tunneling junction. With smaller/larger tip-sample distance $z$, the electric fields at the same tunneling bias are stronger/weaker, so the strength of TIBB effect on the surface resonance states is also stronger/weaker. Such strength of TIBB effect is described by the evolution of the Gaussian peak position in energy with respect to electric field. For convenience, we define the strong/weak TIBB case by Gaussian peaks having different shifting speed with changing tunneling bias (Figure S3), $z$ is set to have a smaller/larger value, respectively.

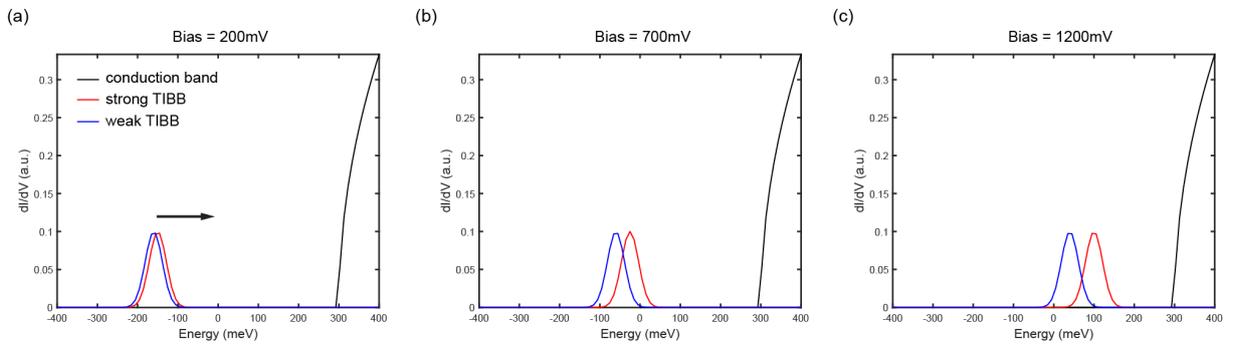

**Figure S3.** Evolution of the Gaussian peak position under different tunneling bias. Black arrow indicates the shifting direction of surface resonance states. Red and blue Gaussian peaks correspond to surface resonance states in strong and weak TIBB cases (the shifting speed is set

to 0.25 meV/1mV and 0.2 meV/1mV), respectively. The tunneling bias is set to be (a) 200mV, (b) 700mV, (c) 1200mV, respectively.

The valence band edge is very close to the Fermi level in BP. So, we define the surface resonance state as entering the energy gap when the associated Gaussian peak crosses the Fermi level (0 meV). In that case, the effective tunneling probability into the bulk are suppressed. We treat this as a negative contribution proportional to the integral of this Gaussian peak above the Fermi level (in the bulk energy gap) to the total conductance.

We calculate tunneling current by integrating the sample density of states, and we find that this negative contribution is due to the suppression of the tunneling probability of surface resonance states in the energy gap. The tunneling current without TIBB is just the integral of the parabolic density of states of BP (Figure S4a). To illustrate the effect of this negative contribution, we calculate the difference of tunneling current with strong/weak TIBB compared to the case without TIBB (Figure S4b). This is done by integrating the part of the Gaussian peak above the Fermi level (with a minus sign since this is a negative contribution). The current drop in the strong TIBB case occurs at a lower bias than the weak TIBB case. This is because with a stronger TIBB effect (corresponding to a smaller tip-sample distance), the electric field in the tunneling junction is sufficiently large to shift the surface resonance states into the energy gap at lower tunneling bias than in the weak TIBB case.

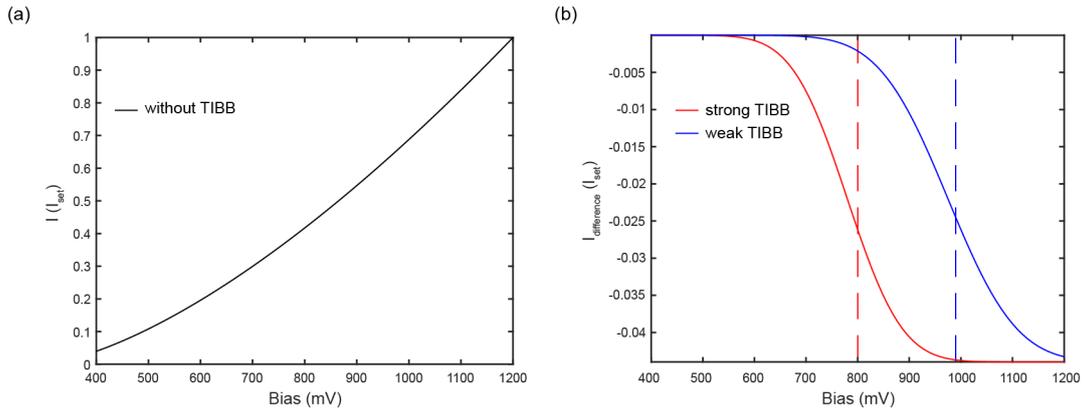

**Figure S4.** Integrated tunneling current and the negative contribution of surface resonance states. (a) Tunneling current calculated by integrating the parabolic conduction band's density of states in Figure S1. (b) Negative contribution of surface resonance states calculated by integrating the part of the Gaussian peak above the Fermi level (with a minus sign). Dashed lines correspond to conductance dip positions. All spectra are normalized by $I_{set} = I$ (Bias = 1200mV).

The differential conductance ($g=dI/dV$) spectra, first derivative of conductance spectra ($dg/dV$), and second derivative of conductance spectra ($d^2g/dV^2$) can then be calculated via doing the corresponding derivative on the calculated total tunneling current. The dip voltage $V_{dip}$ is defined as the peak position in the $d^2g/dV^2$ spectra. This corresponds to the bias at which the peak position of the surface resonance spectral weight is aligned with the Fermi level in each case (800mV and 1000mV for strong and weak TIBB, respectively), resulting in the strongest reduction of the tunneling conductance. The dip position also corresponds to the tunneling bias where the total negative contribution of surface resonance states on tunneling current is increasing the most rapidly (Figure S4b).

The comparison of $g$ and $d^2g/dV^2$ spectra between the simplified model and experimental tunneling data is shown in Figure 3. Figure S5 shows the $dg/dV$ spectra, where a dip-peak feature is centered at the $V_{dip}$ position. It's worth noting that the dip and peak features in $dg/dV$ spectra (~ 0.7 and 0.85 V for strong TIBB, ~ 0.9 and 1.1 V for weak TIBB). Considering the shifting speed of Gaussian peak ($s$ = 0.25 eV/V and 0.2 eV/V, respectively), these bias ranges correspond to an energy shift of ~ 40 meV. This is approximately the FWHM of the Gaussian peak, which corresponds to ~ 70% of the spectral weight from the surface resonance states. So, this bias range corresponds to the whole process of surface resonance states entering the band gap, and in turn marks the range of the full conductance dip feature.

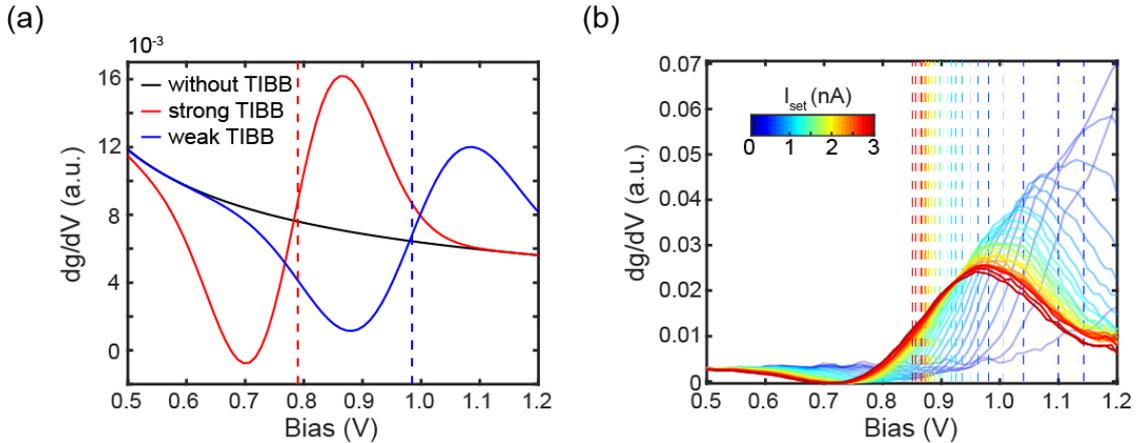

**Figure S5.** Comparison between the simplified model and experimental tunneling spectra. (a) Calculated first derivative of conductance spectra ($dg/dV$) without (black curves), with strong TIBB (red curves) and with weak TIBB (blue curves), respectively. (b) Corresponding experimental $dg/dV$ from Figure 1b measured at different tunneling current setpoints, as indicated by the color scale. Dashed lines indicate the $V_{\text{dip}}$ positions determined by the peak position of $d^2g/dV^2$ spectra in Figures 3b and 3d.

## 5. $V_{\text{dip}}$ evolutions with tunneling junction parameters

As discussed in the main text, when the surface resonance states cross the Fermi level, it shows a crossover from intrinsic semiconducting to doped semiconducting behavior due to the increase in the density of states at the Fermi level. This leads to a reduced TIBB effect on the surface resonance states. Note that in our simplified model, the strength of the TIBB effect corresponds to the shifting rate of the Gaussian peak of the surface resonance state with respect to the tunneling bias. So, the reduction of the TIBB effect can be treated as a reduction of the shifting rate $s$, when the Gaussian peak crosses the Fermi level.

Figure S6 shows the corresponding cases of $z < z_0$ and $z > z_0$. According to our analysis of the effective tunneling barrier (Figure 4), the first case consists of the surface resonance states being "below the valence band edge", "crossing the Fermi level" and "in band gap" (Figure S6a), while the second case only consists of the surface resonance states being "below valence band edge" and "crossing the Fermi level" (Figure S6b), when the tunneling bias is varying from 0 to 1.2V (bias setpoint in our experiments).

When the surface resonance states cross the Fermi level (note that in BP, the valence band edge is located very close to the Fermi level), the shifting rate $s$ of the Gaussian peak is reduced. Figure S6 shows that for $z > z_0$, this reduction affects a broader energy range. So, the total reduction of the shifting rate (or the total TIBB effect, as they are equivalent in our model) is stronger for $z > z_0$. As $V_{\text{dip}}$ measures the strength of total TIBB, this explains the deviation of $V_{\text{dip}}$ in the $V_{\text{dip}}$–$z$ relation toward larger energy position (corresponds to weaker TIBB) for $z > z_0$.

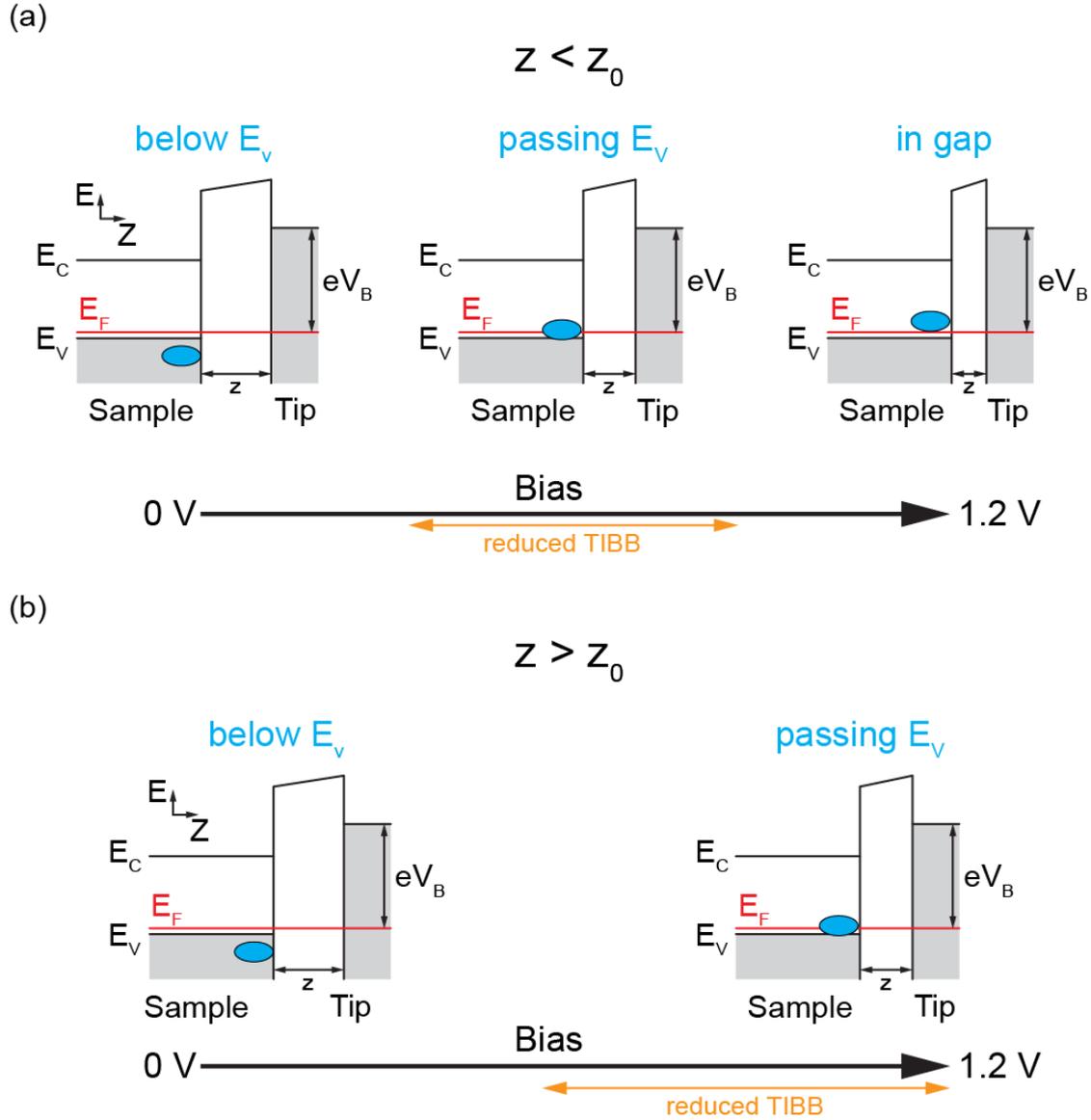

**Figure S6.** Shifting of surface resonance states with varying tunneling bias. (a) For $z < z_0$, the surface resonance states start moving from its original position, crossing the Fermi level and then completely enter the band gap with the bias varying from 0 to 1.2 V. (b) For $z > z_0$, the surface resonance states start moving from its original position, but is still crossing the Fermi level with the bias varying from 0 to 1.2 V. Orange arrows marked the region of the crossing process, where the TIBB effect is reduced.

In contrast to Figure 4, $V_{dip}$ remains linear as a function of the tunneling junction resistance $R_j$ = $V/I_{set}$ remains over the entire experimental range (Figure S7). The $V_{dip}$–$R_j$ relation does not show such a deviation at $z = z_0$. We attribute this to the fact that $I_{set} \propto e^{-2\kappa z}$, where $\kappa$ is the tunneling decay constant ($\varphi$ is included in $\kappa$). Note that the change in $\varphi$ is a result of the surface resonance states contributing to the tunneling conductance when crossing the Fermi level, which corresponds to the reduction of the total TIBB effect. So, $R_j$ already parameterizes the net tunneling transparency—i.e., it implicitly incorporates the exponential dependence on both $z$ and $\varphi$— and thereby remains the same dependency with $V_{\text{dip}}$ over the entire experimental range.

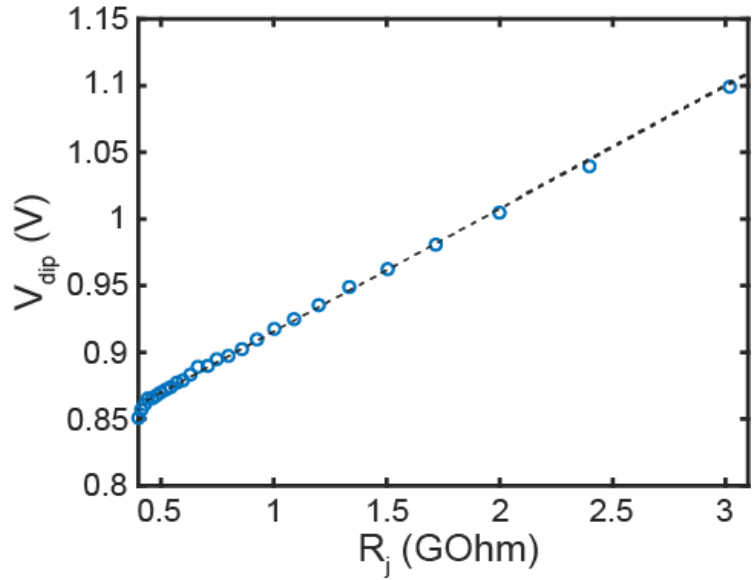

**Figure S7.** $V_{\text{dip}}$ plotted against the tunneling junction resistance $R_{\text{j}} = V/I_{\text{set}}$. Dashed lines show the approximately linear dependence.